\documentclass[tighten,iop]{emulateapj}

\usepackage[english]{babel}
\usepackage{amsmath}
\usepackage{amsfonts}
\usepackage{amssymb}
\usepackage{graphicx}
\usepackage[dvips]{color} 

\usepackage{natbib}

\shorttitle{Jet Collimation in Neutron Star Merger Ejecta}
\shortauthors{Nagakura et al.}

\begin{document}

\title{Jet Collimation in the Ejecta of Double Neutron Star Merger:

 New Canonical Picture of Short Gamma-Ray Bursts  }

\author{Hiroki Nagakura$^{1}$,Kenta Hotokezaka$^{2}$,Yuichiro Sekiguchi$^{1}$,Masaru Shibata$^{1}$, Kunihito Ioka$^{3}$}
\address{$^1$Yukawa Institute for Theoretical Physics, Kyoto
  University, Oiwake-cho, Kitashirakawa, Sakyo-ku, Kyoto, 606-8502,
  Japan}
\address{$^2$Department of Physics, Kyoto University, Kyoto 606-8502, Japan}
\address{$^3$Theory Center, Institute for Particle and Nuclear Studies, KEK, 1-1, Oho, Tsukuba 305-0801, Japan, Department of Particles and Nuclear Physics, the Graduate University for Advanced Studies (Sokendai), 1-1, Oho, Tsukuba 305-0801, Japan}

\begin{abstract}
The observations of jet breaks in the afterglows of short gamma-ray bursts (SGRBs) indicate that the jet has a small opening angle of $\lesssim 10^{\circ}$. The collimation mechanism of the jet is a longstanding theoretical problem. We numerically analyze the jet propagation in the material ejected by double neutron star merger, and demonstrate that if the ejecta mass is $\gtrsim 10^{-2} M_{\odot}$, the jet is well confined by the cocoon and emerges from the ejecta with the required collimation angle. Our results also suggest that there are some populations of choked (failed) SGRBs or low-luminous new types of event. By constructing a model for SGRB 130603B, which is associated with the first kilonova/macronova candidate, we infer that the equation-of-state of neutron stars would be soft enough to provide sufficient ejecta to collimate the jet, if this event was associated with a double neutron star merger.
\end{abstract}


\section{Introduction} \label{sec:intro}

\begin{figure}
\vspace{15mm}
\epsscale{0.9}
\plotone{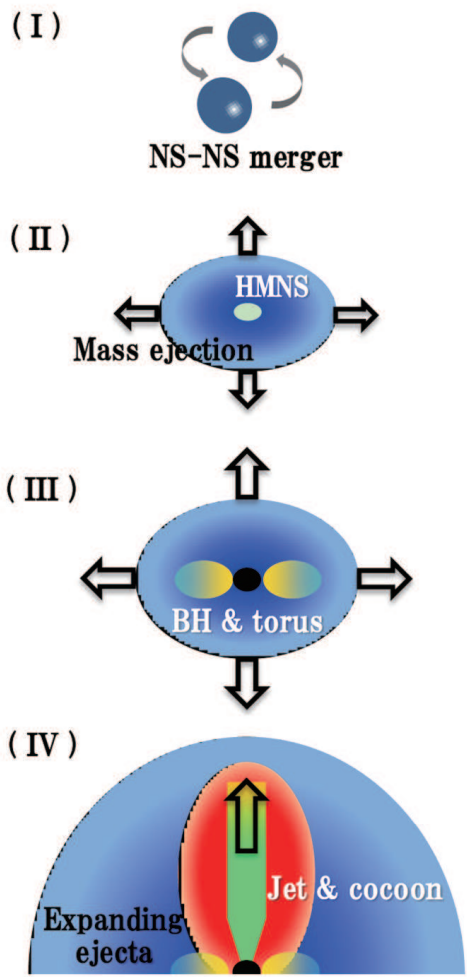}
\caption{The schematic picture of the NS-NS merger scenario for SGRBs. Phase (I): Inspiral phase of NS-NS binary. Phase (II): The mass ejection by the coalescence of NS-NS, and a hypermassive star (HMNS) is formed as a merger remnant, which expels further material from the system. Phase (III): The HMNS collapses to a black hole, and forms the black hole plus torus system. Phase (IV): The central engine starts to operate and the jet propagates through the ejecta.
\label{f1}}
\end{figure}

\begin{table*}
\centering
\caption{\label{tab:model} Models}
\begin{tabular}{lcccccccccc}
\hline\hline
Model~~ &
~$M_{\rm{ej}}$ ($M_{\odot}$)\tablenotemark{a}~  &
~$t_{\rm{i}}$ (ms)\tablenotemark{b}~ &
~$\theta_{0}$ ($^{\circ}$)\tablenotemark{c}~ &
~$L_{\rm{j50}} $\tablenotemark{d}~ &
~$r_{\rm{esc}}$ ($10^{8}$cm)\tablenotemark{e}~ &
~$r_{\rm{max}}$ ($10^{8}$cm)\tablenotemark{f}~ &
~$t_{\rm{b}}$ (ms)\tablenotemark{g}~ &
~$r_{\rm{b}}$ ($10^{9}$cm)\tablenotemark{h}~ &
~$\theta_{\rm{ave}}$ ($^{\circ}$)\tablenotemark{i}~
\\
\hline
{\sl M-ref} &  $10^{-2}$ & 50  & 15 & 2  & 1.2 & 6.1  & 231 & 3.7 & 5.4  \\
{\sl M-L4}      &  $10^{-2}$ & 50  & 15 & 4  & 1.2 & 6.1  & 195 & 3.2 & 5.4 \\
{\sl M-th30}      &  $10^{-2}$ & 50  & 30 & 2  & 1.2 & 6.1  & 626 & 8.9 &5.8 \\
{\sl M-th45}      &  $10^{-2}$ & 50  & 45 & 2  & 1.2 & 6.1  &  -   & -  & - \\
{\sl M-ti500}     &  $10^{-2}$ & 500 & 15 & 2  & 5.6 & 60.1 & 899 & 17.5 & 10.1\\
{\sl M-M3}      &  $10^{-3}$ & 50  & 15 & 2  & 1.2 & 6.1  & 105 & 2.0 & 12.6 \\
{\sl M-M2-2}      & 2 $\times 10^{-2}$ & 50  & 15 & 2  & 1.2 & 6.1  & 320 & 5.0 & 4.7  \\
{\sl M-M1}        &  $10^{-1}$ & 50  & 15 & 2  & 1.2 & 6.1  & 750 & 11.0 & 3.4 \\
\hline\hline
\end{tabular}
\tablecomments{(a) Ejecta mass, (b) Onset timing of jet injection, (c) Initial jet opening angle, (d) Jet power ($L_{\rm{j50}} \equiv L_{\rm{j}}/(10^{50}$erg/s)), (e) Escape radius, (f) Dynamical ejecta front at the time of jet injection, (g) jet breakout time, (h) the radius where the jet head reaches the edge of the ejecta, (i) $\theta_{\rm{ave}}$ at the end of simulations.}
\end{table*}

Recent afterglow observations of short gamma-ray bursts (SGRBs) have provided various information about their environments which
can be interpreted as circumstantial evidence linking SGRBs with mergers 
of compact binaries such as double neutron stars (NS-NS) \citep{1986ApJ...308L..43P,1986ApJ...308L..47G,1989Natur.340..126E} and black hole-neutron star (BH-NS) 
(see \cite{2013arXiv1311.2603B} for a latest review).
On the other hand, the compact binary merger scenario is challenged by the detection of 
jet breaks in the afterglow of some SGRBs and the deduced small jet opening angle 
of $\lesssim 10^{\circ}$ \citep{2006ApJ...650..261S,2006ApJ...653..468B,2011A&A...531L...6N,2012ApJ...756..189F,2013arXiv1309.7479F}. The formation of such a collimated jet in compact binary merger 
has not been clarified yet (see e.g., \citet{2005A&A...436..273A,2012MNRAS.419.1537B}).

One of the most interesting features in the latest numerical-relativity simulations \citep{2013PhRvD..87b4001H} is that NS-NS mergers in general are accompanied 
by a substantial amount of dynamical mass ejection.
Interestingly, the excess in near-IR band observed by {\it Hubble Space Telescope} in {\it Swift} SGRB 130603B 
(\citet{2013Natur.500..547T,2013ApJ...774L..23B})
is explained by the kilonova/macronova model \citep{1998ApJ...507L..59L,2010MNRAS.406.2650M,2013ApJ...774...25K,2013ApJ...775...18B,2013arXiv1307.2943G,2013ApJ...775..113T} 
provided that a large amount of mass 
$\gtrsim 2 \times 10^{-2} M_{\odot}$
is ejected in the NS-NS 
merger and it is powered by the radioactivity of r-process nuclei 
\citep{2013ApJ...778L..16H,2013Natur.500..547T,2014arXiv1401.2166P}.
Such massive ejecta will have a large impact on the dynamics of the jet and
the observed collimation could be naturally explained by their interactions.

In this {\it Letter}, we numerically investigate the jet propagation in 
the material ejected by double neutron star mergers based on a scenario indicated both by our latest numerical-relativity 
simulations and the observations of SGRB 130603B.
The scenario is summarized as follows 
(see Fig.\ref{f1}).

\begin{itemize}

\item According to latest numerical relativity simulations adopting equations of state (EOSs) which are compatible with the recent
discovery of massive neutron stars with $M\sim 2M_{\odot}$ \citep{2010Natur.467.1081D,2013Sci...340..448A}, a hypermassive neutron 
star (HMNS) is the canonical outcome formed after the NS-NS merger for the typical binary mass ($2.6$--$2.8 M_{\odot}$) \citep{2011PhRvL.107e1102S,2013PhRvD..87b4001H,2013ApJ...773...78B}.

\item During and after the merger a large amount of mass $\mathrm{O}(0.01 M_{\odot})$ is ejected (phase (II)) .
This size of ejecta is required to explain the kilonova candidate associated with
SGRB 130603B. According to our numerical-relativity simulations \citep{2013PhRvD..87b4001H}, the morphology of the 
ejecta is quasi spherical for the case of the HMNS formation.
 In particular, the regions along the rotational axis is 
contaminated significantly by the mass ejection.

\item
 Such a large amount of mass
can be ejected only if the EOS of neutron-star matter is relatively soft \citep{2013PhRvD..87b4001H,2013PhRvD..88d4026H,2013ApJ...773...78B}.
In this case, the massive NS formed after the merger is expected to collapse to a BH 
{\it within several tens of milli seconds} (phase (III)), forming a massive torus around it.

\item After the formation of the BH-torus system,
a jet would be launched and it propagates through the expanding merger ejecta (phase (IV)).
A SGRB will be produced only if the jet successfully breaks out of the ejecta.

\end{itemize}

Note that our scenario is different from that explored by previous studies \citep{2005A&A...436..273A} based on
the Newtonian studies \citep{1999A&A...341..499R}, in which the mass ejection is not isotropic but is concentrated along
the orbital plane. In this case, there will be little interaction with the jet and ejecta, and
no collimation by the ejecta is expected. 
Indeed, \citet{2005A&A...436..273A} found no strong collimation by the disk wind (see also \citet{2000PhRvL..85..236L}), since
their simulations were carried out in rather dilute ejecta ($< 10^{-3} M_{\odot}$). 

After studying the dynamics of the jet in the presence of the expanding ejecta, we discuss the canonical model 
for explaining a particular event, SGRB 130603B. With the observationally consistent parameter set, 
we show that relativistic jets successfully break out of the dynamical ejecta and travel with the required collimation angle.

\section{Methods and Models} \label{sec:methodmodel}

\begin{figure*}
 \vspace{15mm}
 \epsscale{0.5}
 \plotone{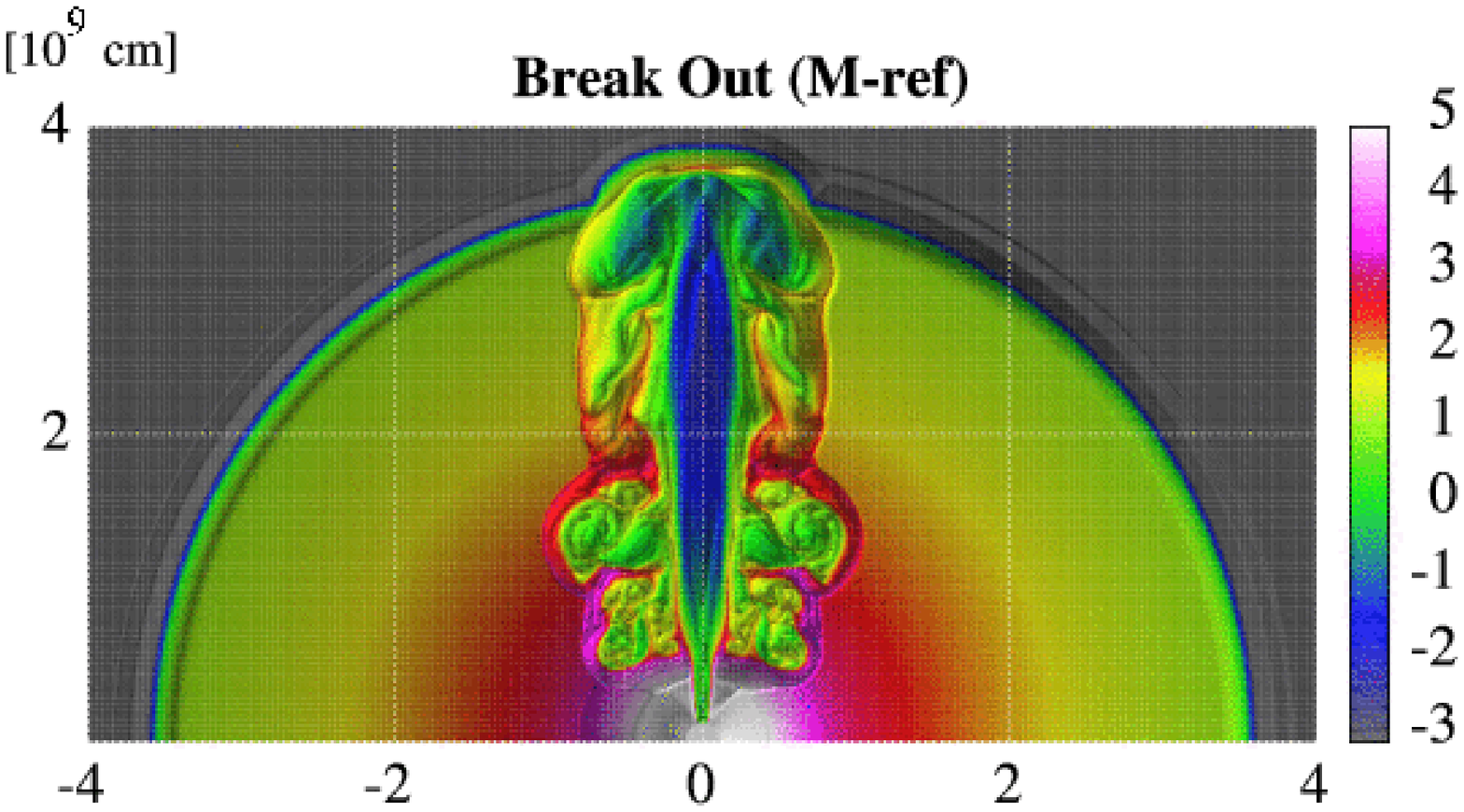}
 \plotone{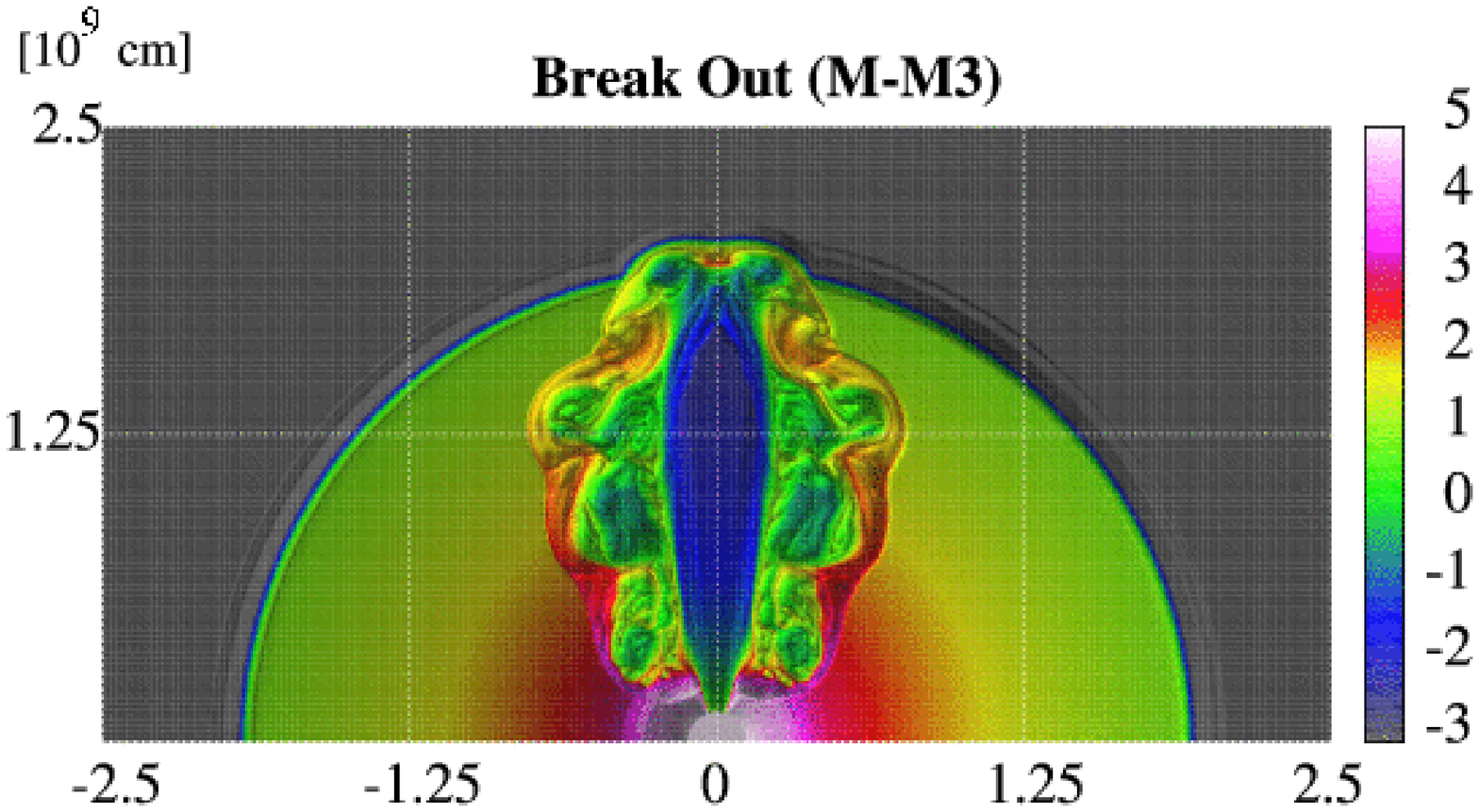}
 \plotone{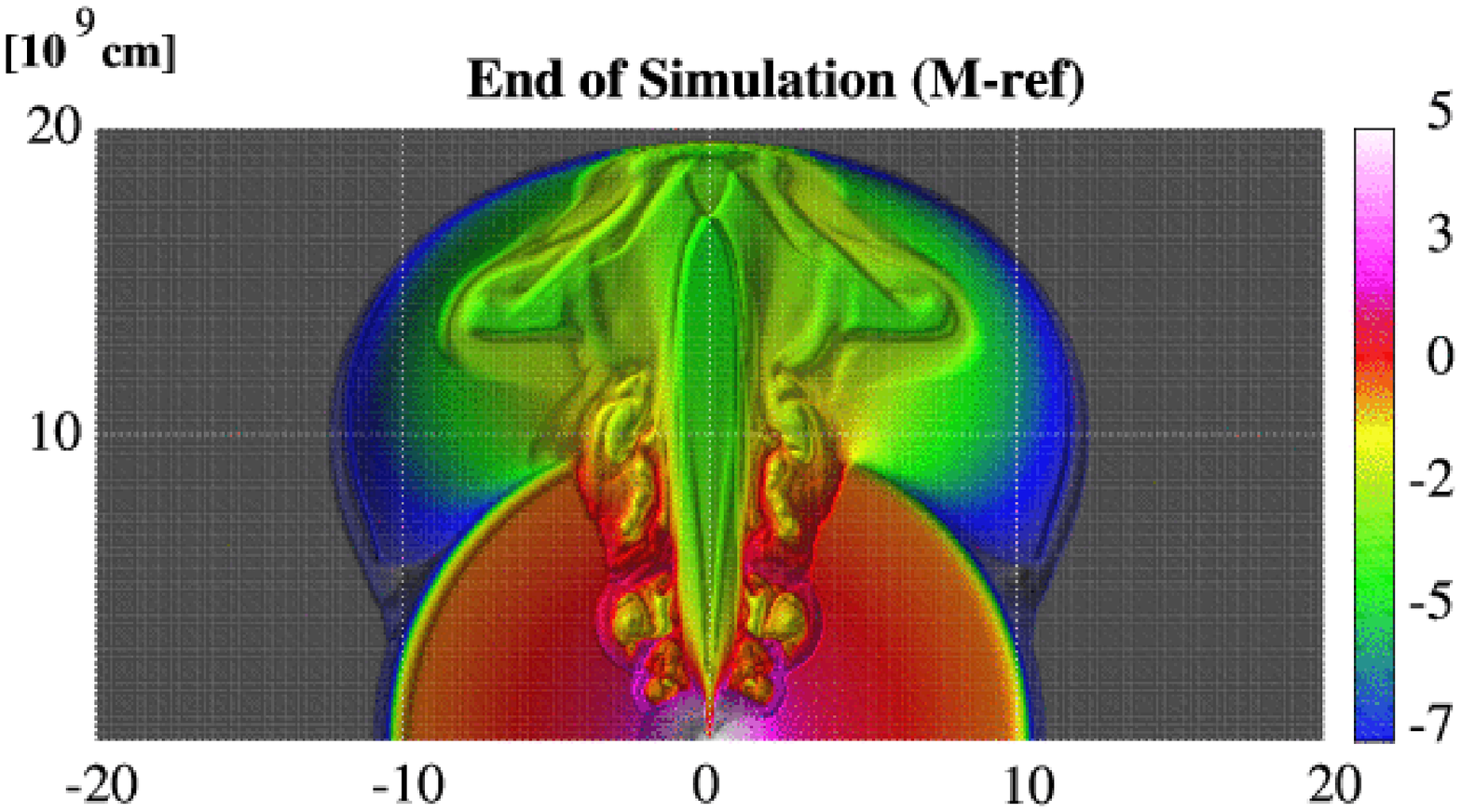}
 \plotone{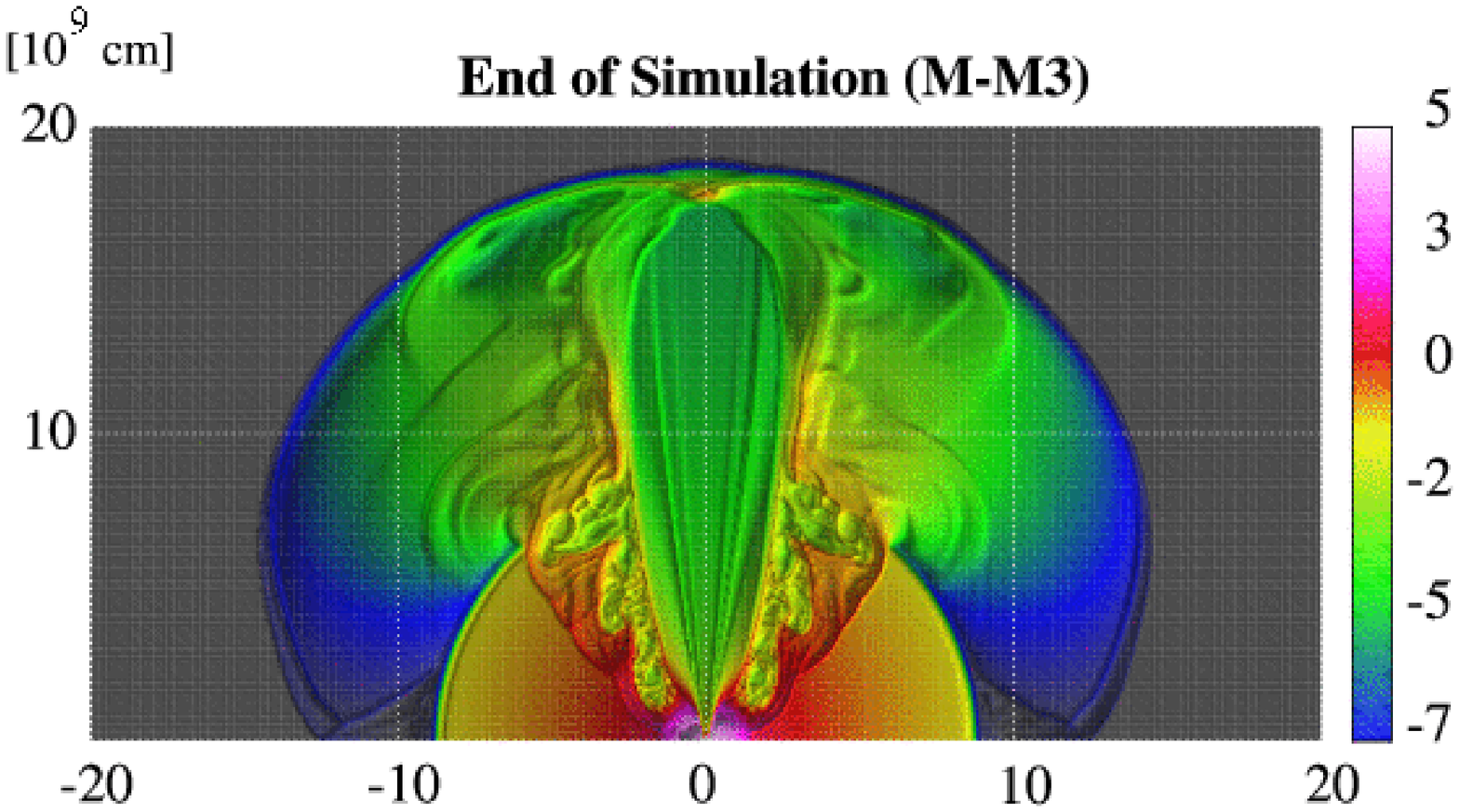}
 \caption{Density contour for two models at $t_{\rm{b}}$ (upper) and the final simulation time (lower). Left:{\sl M-ref}. Right:{\sl M-M3}.
 \label{f2}}
 \end{figure*}

For constructing ejecta profile models, the results from numerical relativity are employed as the reference. We first analyze results in \citet{2013PhRvD..87b4001H}, and then fit the ejecta profile along the pole by the following formulae as
\begin{eqnarray}
&& \rho (t_{\rm{i}},r) = \rho_{0}(t_{\rm{i}}) \biggl( \frac{r}{r_{0}} \biggr)^{-n}, \label{eq:rho} \\
&&r_{\rm{max}} (t_{\rm{i}}) = v_{\rm{max}} (t_{\rm{i}}-t_{0}) + r_{\rm{max0}}, \label{eq:rmax} \\
&& v(t_{\rm{i}},r) = v_{\rm{max}} \biggl( \frac{r}{r_{\rm{max}}} \biggr). \label{eq:velocity}
\end{eqnarray}
In the above expressions, $t_{\rm{i}}$, $r$, $\rho$, and $v$ denote the onset time of jet injection (measured from the merger time), radius, rest-mass density, and velocity of ejecta, respectively. Other variables, $n$, $v_{\rm{max}}$, $r_{0}$, and $t_{0}$ are fitting parameters. The power-law index of density distribution ($n$) has more or less dependence on the dynamics of merger, which is in the range $3 < n < 4$. We choose the middle of this value $n=3.5$ in this study. $v_{\rm{max}}$ denotes
the velocity at the dynamical ejecta front
 (We set $v_{\rm{max}}=0.4c$).
 $t_{0}$ denotes the snapshot time at which we refer to the result of
numerical relativity merger
 simulations.
 We set $t_{0}=10$ms, since the morphology of ejecta has been determined by that time and the outer ejecta continues to be in the homologous expansion phase \citep{2014MNRAS.tmp..181R}.
 The location of forward shock wave at $t_{0}$ is denoted as $r_{\rm{max0}}$, which is set as $r_{\rm{max0}}=1.3 \times 10^{8}$cm. The rest-mass density $\rho_{0}(t_{\rm{i}})$ can be expressed as a function of ejecta mass ($M_{\rm{ej}}$) as;
\begin{eqnarray}
\rho_{0} (t_{\rm{i}}) = \frac{(n-3) M_{\rm{ej}}}{ 4 \pi r_{0}^3 }
 \biggl\{ \biggl( \frac{r_{\rm{esc}}}{r_{0}} \biggr)^{3-n}
  - \biggl( \frac{r_{\rm{max}}}{r_{0}} \biggr)^{3-n}
 \biggr\}^{-1},  \label{eq:rho0}
\end{eqnarray}
where
\begin{eqnarray}
r_{\rm{esc}} = \biggl( \frac{2GM_{\rm{c}} r_{\rm{max}}^2 }{v_{\rm{max}}^{2}}  \biggr)^{\frac{1}{3}}, \label{eq:resc}
\end{eqnarray}
$M_{\rm{c}}$ denotes the central remnant mass, which is chosen as $M_{\rm{c}} \equiv 2.7 M_{\odot}$, and $r_{\rm{esc}}$ denotes the escape radius, which is defined as $v(t_{\rm{i}},r_{\rm{esc}}) \equiv \sqrt{2GM_{\rm{c}}/r_{\rm{esc}}}$. The pressure of ejecta is set as $p = K_{\rm{ef}} \rho ^{4/3}$ with $K_{\rm{ef}}=2.6 \times 10^{15} \rm{g}^{-1/3} \rm{cm}^3 ~ \rm{s}^{-2}$, which is cold enough not to affect the jet and ejecta dynamics.

 According to these formulae, we determine the ejecta profile as a function of $t_{\rm{i}}$ and $M_{\rm{ej}}$.
We first examine the case of $M_{\rm{ej}}= 10^{-2} M_{\odot}$ (see Table~\ref{tab:model}), which is the approximate value of the required mass for explaining the kilonova associated with SGRB 130603B \citep{2013ApJ...778L..16H}, and then we study the dependece on $M_{\rm{ej}}$ ({\sl M-M3, M-M2-2, M-M1}).
 $t_{\rm{i}}$ corresponds to the time of jet injection, which is supposed to be the operation timing of the central engine. For this there are no observational constraints. We set $t_{\rm{i}}=50$ms as the reference value, since our numerical-relativity simulations predict that the life time of HMNS is likely to be several tens of milli seconds to explain the large mass of ejecta $M_{\rm{ej}} \sim 10^{-2} M_{\odot}$ as well as the large mass of torus surrounding a black hole.
 For comparison, we study $t_{\rm{i}}=500$ms case for one model ({\sl M-ti500}, see Table~\ref{tab:model}).

Using the ejecta profile obtained above as initial conditions, we perform axisymmetric simulations of jet propagation by employing a relativistic hydrodynamical code \citep{2011ApJ...731...80N,2012ApJ...754...85N,2013ApJ...764..139N}.
We assume that the central engine successfully operates in the vicinity of the compact remnant, and the jet is injected with constant power from the innermost computational boundary. In these simulations, we focus only on exploring the interaction between the jet and ejecta. Therefore, the computational domain covers from $r_{\rm{esc}}$ to $2 \times 10^{10}$cm. The canonical jet power is set to be $L_{\rm{j}}=2 \times 10^{50}$erg/s for all models, which is comparable with the average jet power of SGRB 130603B (see \citet{2013arXiv1309.7479F} for the collimation-correlated jet energy and also duration of prompt emission). We also prepare the model {\sl M-L4} for which $L_{\rm{j}}=4 \times 10^{50}$erg/s to study the dependece of the jet luminosity.
Throughout our simulations, we use the gamma-law EOS with $\gamma = 4/3$.
The initial Lorentz factor ($\Gamma_{\rm{ini}}$) and specific enthalpy ($h_{\rm{ini}}$) are set to $\Gamma_{\rm{ini}}=5$ and $h_{\rm{ini}}=20$, which result in the terminal Lorentz factor as $\Gamma_{\rm{term}}=100$.
 The initial jet opening angle ($\theta_{0}$) is also not well constrained by observations, and hence we set $\theta_{0}=15^{\circ}$ as the reference value with $\theta_{0}=30^{\circ},45^{\circ}$ for the study of dependence on $\theta_{0}$ ({\sl M-th30, M-th45}). Note that $\theta_{0}=15^{\circ}$ is larger than the opening angle of $1 / \Gamma_{\rm{ini}} = 1/5 \sim 12^{\circ}$, so that the initial thermal expansion of the jet would not be significant (see e.g., \citet{2013ApJ...777..162M}). Simulations are carried out until the shock reaches the outer boundary or time becomes $1$~s after the jet injection. Our models are summarized in Table~\ref{tab:model}.

\section{Jet Dynamics} \label{sec:jetdynamics}

 \begin{figure}
 \vspace{15mm}
 \epsscale{1.2}
 \plotone{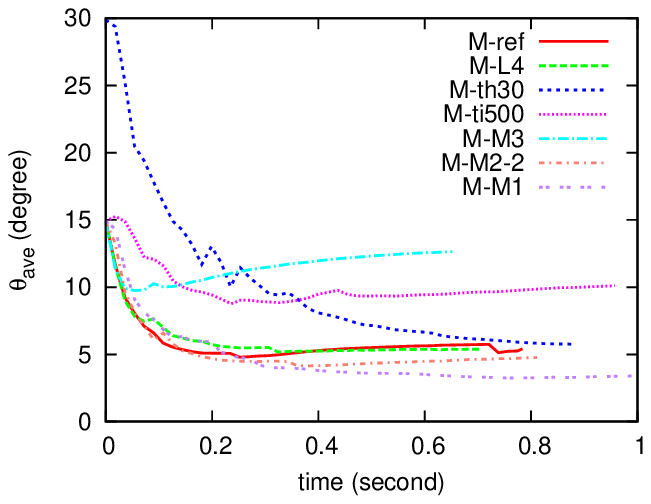}
 \plotone{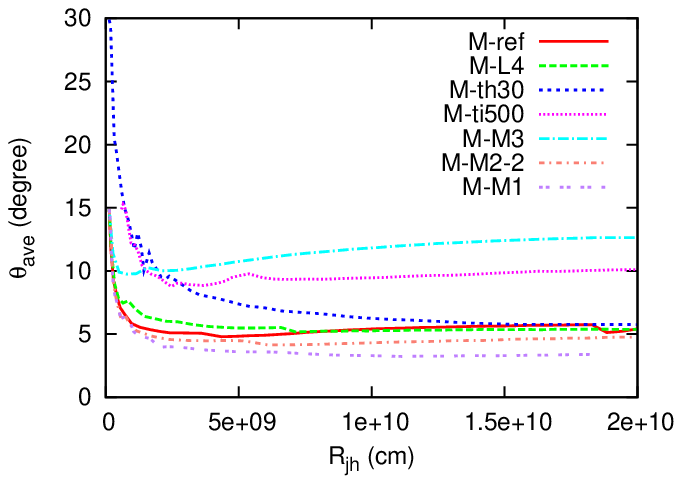}
 \caption{The evolution of average opening angle ($\theta_{\rm{ave}}$) for successful breakout models. Upper: The evolution of $\theta_{\rm{ave}}$ is measured from the time after the jet injection. Lower: Same as the upper one, but the evolution is measured by the location of the jet head ($R_{jh}$).
 \label{f3}}
 \end{figure}

Starting from the initial moment of jet injection at the chosen post-merger time, the jet begins to burrow through the homologously expanding ejecta with mildly relativistic velocity.
 In the left two panels of Fig.~\ref{f2}, we display the density contour maps for {\sl M-ref} at the time of jet breakout and the end of our simulation. At a short distance from the inner boundary, the jet structure changes from conical to cylindrical one due to the confinement by the dense ejecta. The small cross section of the jet head allows the shocked jet matter to escape sideways and generates hot cocoon around the jet. Even though the density gradually decreases with the radius, the surrounding cocoon keeps confining the jet near the pole, and eventually the jet head successfully breaks out of the edge of the ejecta.
 The overall properties of the interaction between ejecta and jet are very similar to those in the context of the collapsar model \citep{2011ApJ...731...80N,2013ApJ...777..162M}. A remarkable difference between the jet propagation in the NS-NS ejecta and the stellar mantle is that the background fluid is no longer stationary and expands with time.
 The jet head chases the ejecta edge from behind, and needs to catch up with it for the relativistic breakout; otherwise it would become non-relativistic ejecta and will never produce SGRBs (see below).

For less massive ejecta case ({\sl M-M3}), the jet experiences less confinement and propagates faster than {\sl M-ref} (see right panels in Fig.~\ref{f2}). Even so, the hot cocoon is formed by the jet-ejecta interaction and works to weakly confine the jet. In order to analyze the cocoon confinement and its degree, we use the dimensionless jet luminosity parameter ($\tilde{L} \equiv \rho_{\rm{j}} h_{\rm{j}} \Gamma_{\rm{j}} / \rho_{\rm{a}}$, where $\rho_{\rm{a}}$ denotes the ambient density above the jet head) following the study by \citet{2011ApJ...740..100B}. By employing equations (\ref{eq:rho})--(\ref{eq:resc}) and imposing the condition $r_{\rm{esc}} \ll r_{\rm{max}}$, $\tilde{L}$ can be roughly estimated as;
\begin{eqnarray}
\tilde{L} \sim  10^{-3} \biggl( \frac{L_{\rm{j}}}{2 \times 10^{50} {\rm{erg/s}}} \biggr)
                          \biggl( \frac{M_{\rm{ej}}}{10^{-2} M_{\odot}} \biggr)^{-1} \nonumber \\
                  \times  \biggl( \frac{\theta_{0}}{15^{\circ}} \biggr) ^{-2}
                          \biggl( \frac{t_{\rm{i}}}{50{\rm{ms}}} \biggr)^{\frac{2}{3}}
                          \biggl( \frac{\epsilon_{r}}{1} \biggr)^{n}
                          \biggl( \frac{\epsilon_{t}}{1} \biggr)^{3-n},
 \label{eq:Ltild}
\end{eqnarray}
where
\begin{eqnarray}
&& \epsilon_{r} \equiv r_{\rm{j}}/r_{\rm{esc}},  \label{eq:epsilonrdef} \\
&& \epsilon_{t} \equiv t/t_{\rm{i}},  \label{eq:epsilontdef}
\end{eqnarray}
and $r_{\rm{j}}$ and $t$ denote the radius of the jet head and the time after the merger, respectively.
According to \citet{2011ApJ...740..100B} \footnote{This criterion is not applicable for the steep density gradient ($n>3$), but we employ it for a qualitative argument. More detailed analytical criterion
 is currently under study \citep{hotoke2014}.}, the condition of cocoon confinement is $\tilde{L} \lesssim \theta_{0}^{-4/3} \sim 6 (\theta_{0}/15^{\circ})^{-4/3}$. In the vicinity of $r_{\rm{esc}}$ ($\epsilon_{r} \sim 1$), all models (including {\sl M-M3}) satisfy the confinement condition, which indicates that the jet undergoes a collimation once at least.
 The cocoon pressure decreases with time because the density of ejecta has steep radial gradient ($n \sim 3.5$). Despite the weakening cocoon pressure, the opening angle of the jet becomes smaller than the initial one.
 In order to analyze the degree of the collimation more precisely, we define the average jet opening angle as
\begin{eqnarray}
\theta_{\rm{ave}}(t) \equiv 
\frac{ \int_{r_{\rm{esc}}}^{R_{\rm{jh}}} \theta_{\rm{op}}(t,r) dr }{ R_{\rm{jh}}(t) - r_{\rm{esc}} },
 \label{eq:thetaave}
\end{eqnarray}
where $R_{\rm{jh}}$ denotes the radius of jet head. The jet opening angle at each radius ($\theta_{\rm{op}}$) is defined as the angle of relativistic components, for which $h \Gamma>10$. Note that if we instead employ the criterion $h \Gamma>100$, we would obtain the incorrectly small $\theta_{\rm{op}}$, caused by baryon pollution by numerical diffusion.
 Figure~\ref{f3} shows the evolution of $\theta_{\rm{ave}}$ for each model. Indeed, $\theta_{\rm{ave}}$ is always less than $\theta_{0}$, which is a clear evidence of a jet collimation. We also find that $\theta_{\rm{ave}}$ after the breakout is larger than $\sim \theta_{0}/5$, which is different from the results in the collapsar case \citep{2013ApJ...777..162M}. This may be attributed to the fact that the ejecta is not stationary contrary to the stellar mantle, and the density gradient of ejecta is steeper than in the case of the stellar mantle.

The initial jet opening angle is also important for the dynamics of jet propagation.
In reality, it would be determined in the vicinity of HMNS or BH by the interaction between the jet and the hot accretion disk \citep{2005A&A...436..273A}, or pinching by magnetic fields
\citep{2006MNRAS.368.1561M}. One of the important consequences of this study is that all models succeed in the breakout by the end of our simulation except for {\sl M-th45} ($\theta_{0}=45^{\circ}$).
For the failed breakout model ({\sl M-th45}),
the shocked jet and ejecta cannot go sideways into the cocoon because of the large cross section of the jet
 and eventually expands quasi-spherically. This fact gives an interesting prediction that there may be some population of choked (failed) SGRBs or low-luminous new types of event, which could be potential candidates for the high energy neutrinos \citep{2001PhRvL..87q1102M,2004PhRvL..93r1101R,2005PhRvL..95f1103A,2008PhRvD..77f3007H,2013PhRvL.111l1102M,2013EPJC...73.2574O}. The rate of these events is uncertain, since it depends on the jet luminosity, opening angle, ejecta mass, and the operation timing of the central engine. We also find that the delayed central engine activity tends to result in failed SGRBs or low-luminous events since the ejecta head has already traveled farther away from the merger remnant (see $r_{\rm{max}}$ of {\sl M-ti500} in Table~\ref{tab:model}).

\section{The canonical model for SGRB 130603B} \label{sec:canoSGRB130603B}

We here discuss the canonical model for SGRB 130603B based on the results of our simulations. According to \citet{2013arXiv1308.2984D,2013arXiv1309.7479F}, SGRB 130603B has a well-collimated jet (its opening angle is $\sim 4 - 8^{\circ}$) with a prompt duration $\Delta T_{90} \sim 200$ms.

Here we focus on the two main properties of the jet: its breakout radius and opening angle. The breakout radius $r_{\rm{b}}$ is defined as the radius where the jet head reaches the edge of the ejecta.
 Broadly speaking, the spatial length of jet ($\Delta l_{\rm{j}}$) in SGRB 130603B is $\Delta T_{90} \times c \sim 6 \times 10^{9}$cm. We regard that $r_{\rm{b}} \lesssim \Delta l_{\rm{j}}$ is a preferred condition for the generation of SGRBs.
In this case, the central engine must be active longer than the jet breakout time $t_{\rm{b}}$, so that the late parts of the jet could reach the emission region without dissipating much energy to the cocoon.
 The duration of the central engine can be estimated as $\Delta t_{ce} \sim t_{\rm{b}} + (\Delta l_{\rm{j}} - r_{\rm{b}})/c$, which is $\sim 300$ms for {\sl M-ref} (see \citet{2012ApJ...749..110B} for a comparison with Long GRBs).
By this criterion, {\sl M-th30}, {\sl M-ti500} and {\sl M-M1} are discarded as the candidate for SGRB 130603B.

The second property we focus on is the jet opening angle and its evolution.
 As shown in the previous section, the jet undergoes the confinement by the ejecta and breaks out with smaller opening angle than the initial one.
The model {\sl M-M3} does not satisfy observational constraints for SGRB 130603B, because the opening angle that it reaches is too large (see Fig.~\ref{f3}).
Therefore, {\sl M-M3} may not be a good model for SGRB 130603B. 
Note that, since $\theta_{\rm{op}}$ includes the jet component inside the ejecta, it is not exactly equal to the observed opening angle. We check the average opening angle of the jet outside of ejecta, and it is not very different from $\theta_{\rm{op}}$.

According to these criteria, {\sl M-ref}, {\sl M-L4} and {\sl M-M2-2} are favored candidates for SGRB 130603B. Note that, if the intrinsic jet luminosity is much larger than $L_{\rm{j}} \sim 10^{50}$erg/s, there is a possibility of production of GRBs even for $\sim 0.1 M_{\odot}$ ejecta mass.
 Note also that if the initial jet opening angle is sufficiently small, it may not require the cocoon confinement to explain the observed small jet opening angle. However, that would become demanding for the central engine, and the mechanism for generating such well-collimated jets has not been discovered yet.

\section{Summary and Discussion} \label{sec:summary}
In this {\it letter}, we investigate the jet propagation in the dynamical ejecta after the NS-NS merger. Similar to the collapsar model, the interaction between the jet and the merger ejecta generates the hot cocoon and
the jet undergoes collimation at least by the deepest and densest layers of the ejecta,
 which is qualitatively consistent with the criterion $\tilde{L} \lesssim \theta_{0}^{-4/3}$. Importantly, models except for quite large initial opening angle ($\theta_{0}=45$) succeed in the breakout with smaller opening angle than the initial one. We also, for the first time, show the possibility that there are some populations for the choked SGRBs or low-luminous new types of event.

Using only the duration of the prompt emission, the jet opening angle, and ejecta mass, we argue for the canonical model for SGRB 130603B.
Under the assumption of spherically symmetric ejecta, {\sl M-M2-2} model satisfies all observational constraints. In reality, however,
 the ejecta profile is not exactly spherically symmetric,
 and its mass contained in the equatorial region tends to be larger.
 According to this,
the ejecta mass in the realistic system would be larger than in our spherical models by a factor of a few.
 Therefore,
 {\sl M-ref} and {\sl M-L4} could also be candidates for SGRB 130603B \citep{2013ApJ...778L..16H,2013Natur.500..547T,2014arXiv1401.2166P}.

 The result of this study and \citet{2013ApJ...778L..16H} suggest that the EOS of neutron stars may be soft among several models of EOS with its maximum mass $> 2 M_{\odot}$ if the central engine of this SGRB is a NS-NS merger.
 The required condition for the central engine is that the jet should be collimated $ \lesssim 15^{\circ}$ before reaching the ejecta, and its life time should be $\sim 300$ms with $L_{\rm{j}} \gtrsim 2 \times 10^{50}$erg/s as the average jet power, and the time lag between merger and jet launching should not be much longer than several tens of milli seconds.

As discussed in this {\it{letter}}, the cocoon confinement changes the conventional picture of jet propagation for the production of SGRBs, and
reinforces the scenario of NS-NS binary merger for SGRBs.
 In BH-NS merger, the morphology of dynamical ejecta is non-spherical, i.e, concentrates on the equatorial plane (see \citet{2013PhRvD..88d1503K}), so the jet never undergoes the strong collimation
unless neutrino or magnetic driven winds from the accretion disk provide enough baryons in the polar region.

\acknowledgments 
We thank Yudai Suwa, Kenta Kiuchi, Kazumi Kahiyama, and Takashi Nakamura for fruitful discussions. This work was supported by Grant-in-Aid for the Scientific Research from the Ministry of Education, Culture, Sports, Science and Technology (MEXT), Japan (22244030, 23740160, 24000004, 24103006, 24244028, 24244036, 24740165, 25103512) and HPCI Strategic Program of Japanese MEXT. The work of K. Hotokezaka is supported by a JSPS fellowship Grant Number 24-1772.

\end{document}